\documentclass[conference]{IEEEtran}
\usepackage{cite}
\usepackage{amsmath,amssymb,amsfonts}
\usepackage{algorithmic}
\usepackage{graphicx}
\usepackage{textcomp}
\usepackage{xcolor}
\usepackage{wrapfig}
\usepackage{comment}
\usepackage{amsthm}
\usepackage{longtable}
\usepackage{ upgreek }
\usepackage{amsmath}
\usepackage{graphicx} 
\usepackage{ gensymb }
\usepackage{ dsfont }
\usepackage{tabularx,booktabs}
\usepackage[acronym,nomain,nonumberlist]{glossaries}
\usepackage{glossary-tree}
\usepackage{tabularx}
\newcolumntype{L}{>{\raggedright\arraybackslash}X}

\usepackage{cleveref}
\usepackage{subcaption}
\usepackage{fancyhdr}

\usepackage[ruled,norelsize]{algorithm2e}
\makeatletter
\newcommand{\removelatexerror}{\let\@latex@error\@gobble}
\makeatother
\SetKw{And}{\hspace{\algoskipindent}\itshape and\;}
\SetKwBlock{Condition}{}{}
\SetKw{Or}{\hspace{\algoskipindent}\itshape or\;}

\makeglossaries
\newacronym{5g}{5G}{Fifth Generation}
\newacronym{mmtc}{mMTC}{massive Machine-Type Com\-mu\-ni\-ca\-tions}
\newacronym{urllc}{URLLC}{Ultra-Reliable Low Latency Communications}
\newacronym{embb}{eMBB}{enhanced Mobile Broadband}
\newacronym{tn}{TN}{Terrestrial Network}
\newacronym{dynasat}{DYNASAT}{"Dynamic spectrum sharing and bandwidth-efficient techniques for high-through\-put MIMO Satellite systems"}
\newacronym{ntn}{NTN}{Non-Terrestrial Network}
\newacronym{mc}{MC}{Multi-Connectivity}
\newacronym{iot}{IoT}{Internet of Things}
\newacronym{vr}{VR}{Virtual Reality}
\newacronym{ngso}{NGSO}{Non-Geostationary Orbit}
\newacronym{dsa}{DSA}{Dynamic Spectrum Allocation}
\newacronym{dss}{DSS}{Dynamic Spectrum Sharing}
\newacronym{poc}{PoC}{Proof-of-Concept}
\newacronym{3gpp}{3GPP}{3rd Generation Partnership Project}
\newacronym{ue}{UE}{User Equipment}
\newacronym{dc}{DC}{Dual Connectivity}
\newacronym{mn}{MN}{Master Node}
\newacronym{sn}{SN}{Secondary Node}
\newacronym{mrdc}{MR-DC}{Multi Radio-Dual Connectivity}
\newacronym{eutra}{E-UTRA}{Evolved Universal Terrestrial Access}
\newacronym{enb}{eNB}{Evolved Node B}
\newacronym{epc}{EPC}{Evolved Packet Core}
\newacronym{5gc}{5GC}{5G Core}
\newacronym{nr}{NR}{New Radio}
\newacronym{engnb}{en-gNB}{en-Next Generation Node B}
\newacronym{ngenb}{ng-eNB}{Next Generation eNB}
\newacronym{ca}{CA}{Carrier Aggregation}
\newacronym{cc}{CC}{Component Carrier}
\newacronym{rrm}{RRM}{Radio Resource Management}
\newacronym{pdcp}{PDCP}{Packet Data Convergence Protocol}
\newacronym{mac}{MAC}{Media Access Control}
\newacronym{rrc}{RRC}{Radio Resource Control}
\newacronym{cn}{CN}{Core Network}
\newacronym{mcg}{MCG}{Master Cell Group}
\newacronym{scg}{SCG}{Secondary Cell Group}
\newacronym{rf}{RF}{Radio Frequency}
\newacronym{ngran}{NG-RAN}{Next Generation Radio Access Network}
\newacronym{gnbcu}{gNB-CU}{gNB-Centralized Unit}
\newacronym{gnbdu}{gNB-DU}{gNB-Distributed Unit}
\newacronym{leo}{LEO}{Low Earth Orbit}
\newacronym{geo}{GEO}{Geostationary Orbit}
\newacronym{ran}{RAN}{Radio Access Network}
\newacronym{hetnet}{HetNet}{Heterogeneous Networks}
\newacronym{rsrp}{RSRP}{Reference Signal Received Power}
\newacronym{ns3}{ns-3}{Network Simulator 3}
\newacronym{gnb}{gNB}{Next Generation Node B}
\newacronym{e2e}{E2E}{End-to-End}
\newacronym{pgw}{PGW}{Packet Network Data Gateway}
\newacronym{sgw}{SGW}{Serving Gateway}
\newacronym{amf}{AMF}{Access and Mobility Management Function}
\newacronym{upf}{UPF}{User Plane Function}
\newacronym{ra}{RA}{Random Access}
\newacronym{4g}{4G}{Fourth Generation}
\newacronym{ap}{AP}{Access Point}
\newacronym{srs}{SRS}{Sounding Reference Signal}
\newacronym{udp}{UDP}{User Datagram Protocol}
\newacronym{sls}{SLS}{System Level Simulator}
\newacronym{kpi}{KPI}{Key Performance Indicator}
\newacronym{ecdf}{eCDF}{empirical Cumulative Distribution Function}
\newacronym{tcp}{TCP}{Transmission Control Protocol}
\newacronym{ahp}{AHP}{Analytic Hierarchy Process}
\newacronym{rat}{RAT}{Radio Access Technology}
\newacronym{sinr}{SINR}{Signal-to-Interference-plus-Noise Ratio}
\newacronym{ts}{TS}{Technical Specification}
\newacronym{tr}{TR}{Technical Report}
\newacronym{lan}{LAN}{Local Area Network}
\newacronym{rnti}{RNTI}{Radio Network Temporary Identifier}
\newacronym{sib}{SIB}{System Information Block}
\newacronym{mib}{MIB}{Master Information Block}
\newacronym{nlos}{NLOS}{Non-Line of Sight}
\newacronym{rng}{RNG}{Random Number Generator}
\newacronym{sue}{SUE}{Spectral Utilization Efficiency}
\newacronym{rb}{RB}{Resource Block}
\newacronym{re}{RE}{Resource Element}
\newacronym{ewma}{EWMA}{Exponential Weighted Moving Average}
\newacronym{sdap}{SDAP}{Service Data Adaption Protocol}
\newacronym{ho}{HO}{Handover}
\newacronym{pdu}{PDU}{Protocol Data Unit}
\newacronym{wa}{WA}{Wraparound}
\newacronym{cbr}{CBR}{Constant Bit Rate}

\def\BibTeX{{\rm B\kern-.05em{\sc i\kern-.025em b}\kern-.08em
    T\kern-.1667em\lower.7ex\hbox{E}\kern-.125emX}}

\renewcommand\IEEEkeywordsname{Keywords}

\makeatletter
\newcommand{\linebreakand}{%
  \end{@IEEEauthorhalign}
  \hfill\mbox{}\par
  \mbox{}\hfill\begin{@IEEEauthorhalign}
}
\makeatother

\fancypagestyle{FirstPage}{

\lfoot{\copyright 2023 IEEE. Personal use of this material is permitted. Permission from IEEE must be obtained for all other uses, in any current or future media, including reprinting/republishing this material for advertising or promotional purposes, creating new collective works, for resale or redistribution to servers or lists, or reuse of any copyrighted component of this work in other works. DOI 10.1109/WoWMoM57956.2023.00074}

}

\begin{document}
\bstctlcite{IEEEexample:BSTcontrol}
\newtheorem{thm}{Theorem} 
\theoremstyle{definition}
\newtheorem{remark}[thm]{Remark}
\newtheorem{defn}[thm]{Definition}
\theoremstyle{plain}
\newtheorem{thr}[thm]{Theorem}
\newtheorem{prop}[thm]{Proposition}
\newtheorem{kor}[thm]{Corollary}

\title{Coordinated Dynamic Spectrum Sharing Between Terrestrial and Non-Terrestrial Networks in 5G and Beyond}

\author{
\IEEEauthorblockN{Henrik Martikainen\IEEEauthorrefmark{1}, Mikko Majamaa\IEEEauthorrefmark{1}\IEEEauthorrefmark{2}, Jani Puttonen\IEEEauthorrefmark{1}}

\IEEEauthorblockA{
\IEEEauthorrefmark{1}\textit{Magister Solutions Ltd, Jyv\"{a}skyl\"{a}, Finland} \\
email: \{firstname.lastname\}@magister.fi
}

\IEEEauthorblockA{
\IEEEauthorrefmark{2}\textit{Faculty of Information Technology, University of Jyv\"{a}skyl\"{a}, Jyv\"{a}skyl\"{a}, Finland}}

}

\maketitle

\begin{abstract}
The emerging Non-Terrestrial Networks (NTNs) can aid to provide 5G and beyond services everywhere and anytime. However, the vast emergence of NTN systems will introduce an unseen interference to both the existing satellite systems and Terrestrial Networks (TNs). For that, there is a need for novel ideas on how to efficiently utilize the co-existing systems with the ever-increasing competition on scarce spectrum resources. Dynamic Spectrum Sharing (DSS) is a promising technique in which different systems can operate on the same spectrum, thus increasing the spectrum efficiency and offering better coverage for the users. In this paper, we present a centralized scheme for achieving coordinated DSS to protect the primary TN while providing NTN with sufficient resources. The scheme is evaluated by system simulations in a scenario with a TN and low earth orbit satellite. The results reveal that in a low traffic demand situation, the primary TN users are not affected negatively while the NTN can provide service to the rural area. In high-demand traffic situations, the peak performance of the TN inevitably suffers but the TN cell edge and NTN users' performance is improved.
\end{abstract}


\section{Introduction}
\label{sec:introduction}

\thispagestyle{FirstPage}

New Radio (NR), the air interface of 5G, is the first 3GPP mobile communications standard that supports Non-Terrestrial Networks (NTNs) communications from the go. The NTN standardization in 3GPP started in its Release 15 and 16 with study items for NR to support NTNs. Release 17, finalized in 2022, included basic functionalities to enable NR for NTNs. Release 18 marks the beginning of standardization toward the 5G-Advancend (5G-A) and 6G. NTN-wise this means, for example, expanding coverage to higher frequencies and considering mobility issues.

NTNs have attracted a lot of attention from the industry and academia in recent years. The cost of such systems has gone down, which has attracted new players to the pool of satellite communication providers.
Especially Non-Geostationary Orbit (NGSO) satellite systems have gained a lot of attention due to their relatively cheap price, deployment cost, and shorter propagation delays (in an order of magnitude compared to the traditional GSO satellites). The shorter propagation delays make them suitable for a plethora of applications that are impractical when GSO satellites are involved.

However, the vast emergence of NTN systems will introduce an unseen interference to both the existing satellite systems and Terrestrial Networks (TNs). For that, there is a need for novel ideas on how to efficiently utilize the co-existing systems with the ever-increasing competition on scarce spectrum resources. Dynamic Spectrum Sharing (DSS) is a promising technique in which different systems can operate on the same spectrum, thus increasing the Spectrum Efficiency (SE) and offering better coverage for the users.

The different TN/NTN systems can operate either on a licensed or unlicensed spectrum. When operating on a licensed spectrum, the whole spectrum is primarily reserved for a single system whereas an unlicensed spectrum may be utilized by any system. Licensed spectrum is typically auctioned by the local spectrum regulatory authority, e.g., Federal Communications Commission (FCC) in the United States. Depending on the spectrum type, i.e., licensed/unlicensed, different DSS techniques can be utilized. For example, for the former, Licensed Spectrum Access (LSA) can be utilized in which a Primary User (PU) of the spectrum is the incumbent user of the spectrum, but Secondary Users (SUs) can access the spectrum when it is available. The availability of the spectrum can be deduced from a spectrum utilization database, or it can be measured with spectrum sensing techniques. Further, Concurrent Spectrum Access (CSA) can be utilized in which the SUs may transmit concurrently with the PU with limitations on transmit power. DSS for unlicensed spectrum can be achieved, e.g., by Listen Before Talk (LBT) mechanisms as in Wi-Fi.

Next, related literature is briefly surveyed. A DSS method between LEO and GEO satellites is introduced in \cite{ 9163369}. In the scheme, one LEO satellite senses the spectrum while another data LEO satellite transmits based on the measurements. However, there may be difficulties and uncertainty associated with spectrum sensing approaches \cite{ 9910435}. The authors in \cite{ 8101462} survey database-assisted spectrum sharing in satellite communications. One of the potential spectrum sharing scenarios is identified as NTN as an SU of the spectrum. One of the problems identified is how to consider all the relevant information in spectrum allocations while protecting the PUs from interference and still offering enough capacity to SUs. In \cite{ 8793059}, a centralized DSS scheme (for TNs) is proposed in which a central entity computes interference graphs and based on them, signals spectrum allocations to the base stations.

As the standardization toward 6G progresses it is expected that the cooperation between the TN and NTN operators will increase. Even though many successful schemes for DSS have been proposed, there is a need for research on schemes in which tight cooperation to share the spectrum between a TN and NTN can be exploited. In this paper, we extend the idea of a centralized entity that is responsible for spectrum allocations to a scenario in which a satellite without the assumption of sensing capabilities is involved. Further, instead of computing the allocations based on interference, the allocations are adjusted based on load metrics. The TN and NTN are considered respectively the PU and SU of the spectrum and Coordinated Dynamic Spectrum Sharing (C-DSS) is performed between them. The scheme is evaluated with a cutting-edge 5G NTN System-Level Simulator (SLS) with packet-level precision.

The rest of the paper is organized as follows. In Section~\ref{sec:dssbtantn}, DSS between TN and NTN is discussed and the proposed framework is introduced. In Section~\ref{sec:simulations}, the framework is evaluated through comprehensive system-level simulations. Finally, in Section \ref{sec:conclusions}, the concluding remarks are provided.

\section{Dynamic Spectrum Sharing Between Terrestrial and Non-Terrestrial Networks}
\label{sec:dssbtantn}

\subsection{Coordinated Dynamic Spectrum Sharing Basic Principle}

\begin{figure}[htb!]
    \centering
    \includegraphics[width=\linewidth]{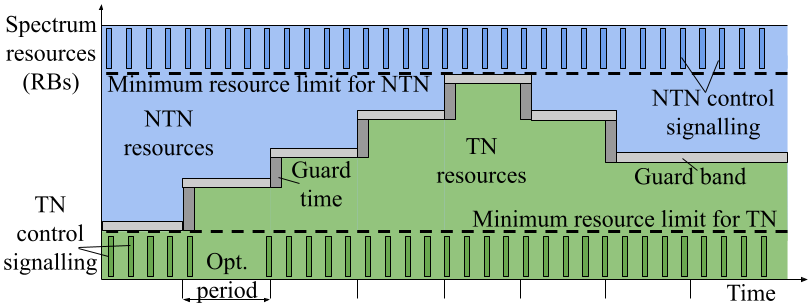}
    \caption{Radio resource usage for TN and NTN with C-DSS.}
    \label{fig:reservation}
\end{figure}

5G New Radio (NR) frame structure is very flexible, and it has several options for coordinating transmissions of different network entities. The implementation of TN-NTN coordinated spectrum sharing in this paper is based on changing the data resource blocks (RBs) allocation in the gNBs using the existing 3GPP specifications. During the RB allocations assignment, a range of RBs is reserved as a guard band. Guard bands are a narrow frequency range (RBs) left unused to prevent interference between the TN and NTN simultaneous transmission. The resource allocation considers guard bands and guard intervals as shown in Fig.~\ref{fig:reservation}. While the networks are assigned to a shared bandwidth, each operates on a limited range of RBs.

{}

\subsection{Proposed Coordinated Dynamic Spectrum Sharing Architecture}

\begin{figure}[htb!]
    \centering
    \includegraphics[width=\linewidth]{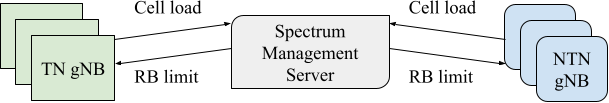}
    \caption{Proposed SMS architecture.}
    \label{fig:architecture}
\end{figure}

In the C-DSS architecture here, a centralized approach is implemented to manage the coordinated spectrum sharing between the coordinated sharing parties which are the TN and NTN operators in this case. The data resource block allocation for each network will be decided by the allocation algorithm in the Spectrum Management Server (SMS) system. The SMS system is an entity that receives load information as an input and sets spectrum limits for each network as an output. Based on the received input network load metrics the SMS calculates the total network load for each network and signals the assigned RB range for the TN and NTN. The load information affects the range of RB resources to be allocated for the TN and NTN. The SMS architecture is presented in Fig.~\ref{fig:architecture}.

{}

\subsection{Coordinated Dynamic Spectrum Sharing Algorithm}

The C-DSS algorithm helps to define the optimal spectrum sharing between the TN and NTN. The dynamic allocation varies between these minimum capacity limits depending on the time-varying loads the networks experience. To make an optimal resource-sharing decision, the algorithm flows steps are initialization, load evaluation, and resource allocation update. The TN is given higher priority and if there are unutilized TN resources those are given to the NTN. In addition, the algorithm considers the guard bands needed between the systems.

During the initialization process, the algorithm collects information about the initial RB values of each system. The initial RB values will be verified to obey the minimum resource reservation. The algorithm ensures that at least the minimum amount of resources are reserved for each system. Fig.~\ref{fig:reservation} illustrates the minimum reservation and the available resources for dynamic allocation.

We have assumed that all the transmissions by both systems always obey fully the coordinated spectrum resources, i.e., also all the control messaging happens on the fixed spectrum portion of each system. This way there is no need to synchronize the spectrum usage in the time domain and the guard time is only needed for those few updated RBs. The guard time is needed since the TN and NTN are not synchronized. The details on how the control messaging is implemented on the limited fixed spectrum portion, if any specification updates are needed and how it affects the system performance, are left for future studies.

In the load evaluation step, the algorithm calculates the average traffic load of the TN from individual TN cell reports. After the load average  has been calculated, the algorithm observes where the values lie in comparison to configurable lower and upper threshold values for TN load.  If the load of the TN is greater than the upper threshold value, the system is considered in need of more resources. In this case, the algorithm increases the RB ranges of the system by a configured amount. In the case that the load of the system lies in between the threshold values, there is no update needed to the RB values of the systems. On the other hand, if the data load is less than the lower threshold value, a range of RBs is given to the NTN system. The algorithm keeps doing this system load evaluation procedure periodically, e.g., every 30 seconds. In the last step, the algorithm updates the resource allocations and signals them to both systems.

\begin{figure}[htb!]
    \centering
    \includegraphics[width=\linewidth]{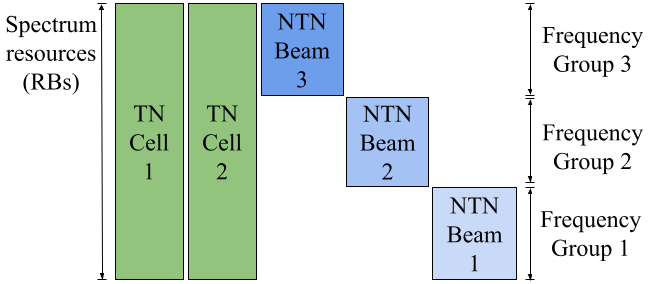}
    \caption{Frequency groups.}
    \label{fig:cdss_groups}
\end{figure}

The above explanation assumes that the TN and NTN systems defined are fully overlapping in frequency which is not always the case. Typically, TNs  use a Frequency Reuse Factor 1 (FRF1) while satellite systems must use, e.g., FRF3 due to less defined cell areas formed by the satellite beams, especially when neighboring cells are formed by beams from the same satellite i.e., a multi-beam satellite system is used like here. Fig.~\ref{fig:cdss_groups} shows the frequency band sharing between TN and NTN in the case of FRF1 and FRF3, respectively. In this case, the frequency band is divided into three frequency groups and each NTN beam in the system belongs to one of them, and the TN bandwidth is divided into these three groups. In order not to have conflicting decisions, the implemented C-DSS algorithm only analyzes and adjusts the resources of one frequency group at a time.

{}

In addition to the above frequency dimension correlation of resources of the two systems, there needs to be a spatial dimension correlation. Clearly, TN and NTN cells that are far away from each other are not interfering each other while overlapping cells using the same RBs are. The implemented C-DSS algorithm provides a simple framework for that by having an option of manually disabling the coordination of some of the frequency groups, i.e., RBs. All the remaining RBs are fully coordinated, for example, if frequency group 3 coordination is disabled, those RBs can be fully utilized by all TN and NTN cells. While RBs within groups 1 and 2 are only utilized by either TN or NTN, but not both. In larger scenarios, there needs to be a way to group the TN and NTN cells which interfere with each other into the same coordination groups. However, in the presented simulation scenario, only a single coordination group is considered.

The simulation optimization period is very short at 0.25 s which is too short an optimization period for a real deployment. This was used to achieve a converged RB share situation during each simulation run while still utilizing the same serving satellite. Also, due to the constant traffic and stable nature of the simulation users, there is no need to have a longer period. Still, in real life, the optimization period must be longer. A LEO 600 satellite moves at a speed of 7.56 km/s and with Earth moving beams the NTN serving areas are constantly changing and the proposed C-DSS method is very hard to implement. However, with quasi-Earth fixed beams the serving area of a beam remains similar, and a single satellite serves the beam significantly longer, up to a few minutes, depending on the constellation used. This is unlike Earth fixed beams of geostationary satellites which never change the pointing. Also, there is no need to reach a converged state during a single satellite service time, but the optimization state can be associated with the cell area on the ground and optimization can be done over multiple satellite service times. Thus, we believe that similar results can be expected with the proposed C-DSS approach in a longer scenario with more dynamic traffic.

\section{Simulations}
\label{sec:simulations}

\subsection{5G Non-Terrestrial Network Simulator}

The introduced C-DSS implementation is evaluated by system simulations. The 5G NTN SLS \cite{ntn} used in the study is based on Network Simulator 3 (ns-3) \cite{Riley2010} and its 5G LENA module \cite{Patriciello2019AnES}. Ns-3 is primarily utilized for educational and research purposes as a discrete-event non-real-time packet-level network simulator that can be expanded with new modules. One such module is 5G LENA which is designed for simulating 5G networks. However, it cannot simulate NTNs. To enable the simulation of NTNs, the necessary components were added to 5G LENA, which served as a starting point for the development of the 5G NTN SLS.

The 5G LENA module incorporates NR Physical (PHY) and Medium Access Control (MAC) features, while the upper layers of the UE/gNB stack are borrowed from the ns-3 Long Term Evolution (LTE) module \cite{lenalte}. The link layer is abstracted using a Link-to-System (L2S) mapper and Modulation and Coding (MODCOD)-specific Signal-to-Interference plus Noise Ratio (SINR) to Block Error Rate (BLER) mapping curves. For each packet, SINR is calculated and then, with the help of the mapper, BLER is inferred. Further, the higher layers such as transport and network layers are provided by the ns-3 Internet model.

The 5G NTN SLS has been used extensively in the past in Research \& Development (R\&D) efforts, e.g., in the European Space Agency (ESA) ALIX project \cite{alix} aiming to support 5G NTN standardization in 3GPP, and in the Dynamic Spectrum Sharing and Bandwidth-Efficient Techniques for High-Throughput MIMO Satellite Systems (DYNASAT) project \cite{dynasat} to simulate bandwidth-efficient and DSS techniques. 

The simulator has been calibrated as part of previous R\&D efforts, utilizing system-level calibration results from \cite{38821}. Additionally, channel and antenna/beam modeling from \cite{38811} have been incorporated, along with a global coordinate system and calibration scenarios from [11]. The calibration scenarios serve as a baseline for parametrization and can be adjusted according to specific needs. These scenarios make it possible to research various assumptions such as different bands (S-band/KA-band), terminal types (VSAT, handheld), and frequency reuse patterns (reuse 1, 3, 2+2), and allow for the study of hybrid TN-NTN scenarios.

Regarding the C-DSS implementation in the simulator, currently, it is assumed that the SMS is a floating object that communicates with the TN and NTN with ideal callbacks. The signaling between the SMS and TN/NTN system is implemented as function calls to get updated input/output information to/from the SMS. The details of real communication protocol are left for future study. 

\subsection{Scenario and Assumptions}

The scenario aims to show how a user with an NTN-enabled mobile terminal could benefit from NTN and C-DSS. The main use case for NTN is coverage extension in regions where TN cannot be deployed, or the deployment is too expensive, e.g., mountain or sea areas. To minimize the signal attenuation a low carrier frequency of S-Band, i.e., 2 GHz is the most attractive option. However, the available frequency bands for S-Band are very limited and are mostly already reserved for TN operators. Thus, it is likely that the same band is shared between TN and NTN operators. Here we study C-DSS in a scenario where the NTN operator wants to maximize the NTN coverage and thus the NTN beams are operating close to the existing TN. At the same time, we assume that the TN and NTN operators cooperate and have a complete operating C-DSS framework along with an SMS.

\begin{figure}[htb!]
    \centering
    \includegraphics[width=\linewidth]{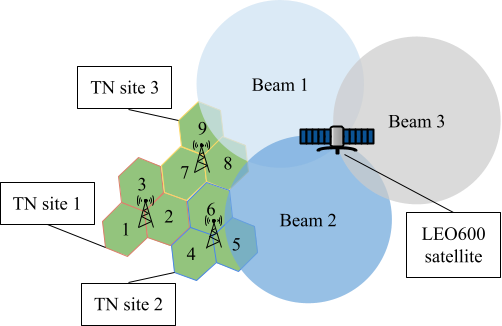}
    \caption{Simulation scenario.}
    \label{fig:scenario}
\end{figure}

Fig.~\ref{fig:scenario} shows the simulation scenario layout. The simulation scenario consists of one 5G NTN LEO satellite located at 600 km altitude and a 5G TN consisting of 3 sites each with 3 sectors (9 cells in total). The Inter-Site Distance (ISD) of TN cells is 7500 m. LEO satellite is configured with three beams using Set-1 parameters from \cite{38821}. NTN 3dB beam radius is 25 km. The satellite is moving and quasi-Earth-fixed configuration is used. The elevation angle during the simulation varies from 70 to 80 degrees (90 degrees is fully vertical) which does not have a significant impact on the beam pattern on the ground. The simulation time is 10 s and all the results are collected from the final 5 s which is enough for the C-DSS algorithm to reach a converged state.

Five UEs are placed within each NTN beam coverage area and ten users are placed in each TN cell area bringing the total UE count to 105. UEs are allowed to make cell selection autonomously resulting that UEs being connected to the best possible beam or cell; some of the UEs are connected to the NTN beam, while some of the UEs are connected to one of the TN cells. Traffic is either 400 or 1200/4000 Kbps constant bitrate over UDP in the downlink (DL) direction. A comprehensive list of simulation parameters is listed in Table~\ref{table:params}.

{}

\begin{table}[]
\caption{Simulation parameters.}
\label{table:params}
\begin{tabularx}{\linewidth}{l|L}
\hline
\textbf{Parameter}               & \textbf{Value}                                   \\ \hline
Simulation Time                  & 10.0 s                                            \\
Warmup Time                      & 5 s                                            \\ 
Satellite Mobility               & Moving                                           \\ 
UE Mobility                      & Stationary                                       \\ 
Beam Deployment                  & Quasi-Earth Fixed                                \\ 
NTN Channel Condition            & LOS                                              \\ 
TN Channel Condition             & Dynamic LOS                                     \\ 
Number of TN cells               & 9                                               \\ 
Number of NTN beams              & 3                                                \\ 
UEs per TN cell area             & 10                                               \\ 
UEs per NTN beam area            & 5                                               \\
Total UE count                   & 105                                               \\
ISD                              & 7.5 km                                           \\ 
TN Deployment                    & Rural                                            \\ 
Bandwidth per NTN beam           & 10 MHz                                           \\ 
Bandwidth per TN sector          & 30 MHz                                           \\ 
NTN carrier frequency            & 2 GHz (S-band)                                   \\ 
Satellite Orbit                  & LEO 600 km                                       \\ 
Satellite Parameter Set          & Set 1 \cite[Table 6.1.1.1-1]{38821}                   \\ 
UE Antenna Type                  & Handheld omni-directional 0dBi gain\\ 
Traffic                     & CBR with UDP                                              \\ 
Communication Protocol           & UDP                                              \\ 
Traffic Demand per UE                  & 400/1200/4000 Kbps                           \\ 
HARQ                             & Enabled                                          \\ 
Min number of RB per cell           & 12 \\
RNG Runs                         & 10  \\
\hline
\textbf{C-DSS algorithm parameters}\\
\hline
Min resources for TN/NTN & 6/6 RBs\\
Target load for TN & 60-80\% \\
Guard bands & 3 RBs\\
C-DSS opt. period & 0.25s\\
\hline
\end{tabularx}
\end{table}

Table~\ref{table:simcases} lists the simulation cases. There are two traffic cases, 400 Kbps (TN and NTN) and 4/1.2 Mbps (TN/NTN), designated as Low-Demand (LD) and High-Demand (HR), respectively. Additionally, there are two spectrum configurations. In the "TN only" case NTN beam is completely disabled and TN can utilize the full 30 MHz bandwidth. The NTN beam share is 78\% and 48\% with C-DSS LD and HD traffic cases, respectively. In the "C-DSS" cases the C-DSS algorithm tries to achieve 60-80 \% TN load by adjusting the available resources and presented shares are the outcome after the C-DSS optimization is finalized. In these C-DSS cases, the NTN beam 3 spectrum, i.e., frequency group 3, is not coordinated but is fully usable both for TN and NTN. Due to the large distance between beam 3 and the TN, it can be assumed that there is no inter-system interference. This could be monitored by the SMS using interference measurements from TN cells and the decision for disabling the coordination can be done. The research and the implementation of this aspect are left for future studies.

\begin{table}[]
\caption{Simulation cases.}
\label{table:simcases}
\begin{tabularx}{\linewidth}{l|L|X|X}
\hline
\textbf{Case} & \textbf{TN cell BW (out of 30MHz)} & \textbf{NTN beam BW (out of 10MHz)} & \textbf{Traffic demand per TN/NTN UE} \\ \hline
1: TN only LD& 100 \% & 0 \% & 400/400 Kbps\\
2: C-DSS LD& 47 \% & 78 \% & 400/400 Kbps\\
3: TN only HD& 100 \% & 0 \% & 4/1.2 Mbps\\
4: C-DSS HD& 76 \% & 48 \% & 4/1.2 Mbps\\
\hline
\end{tabularx}
\end{table}

\subsection{Results}

\begin{figure}[htb!]
    \centering
    \includegraphics[width=.75\linewidth]{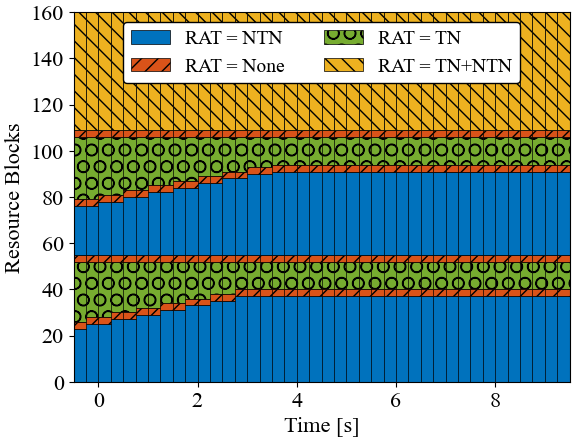}
    \caption{C-DSS resource allocation during case 2 C-DSS LD simulation.}
    \label{fig:cdsa_rbs}
\end{figure}

\begin{figure}[htb!]
    \centering
    \includegraphics[width=.75\linewidth]{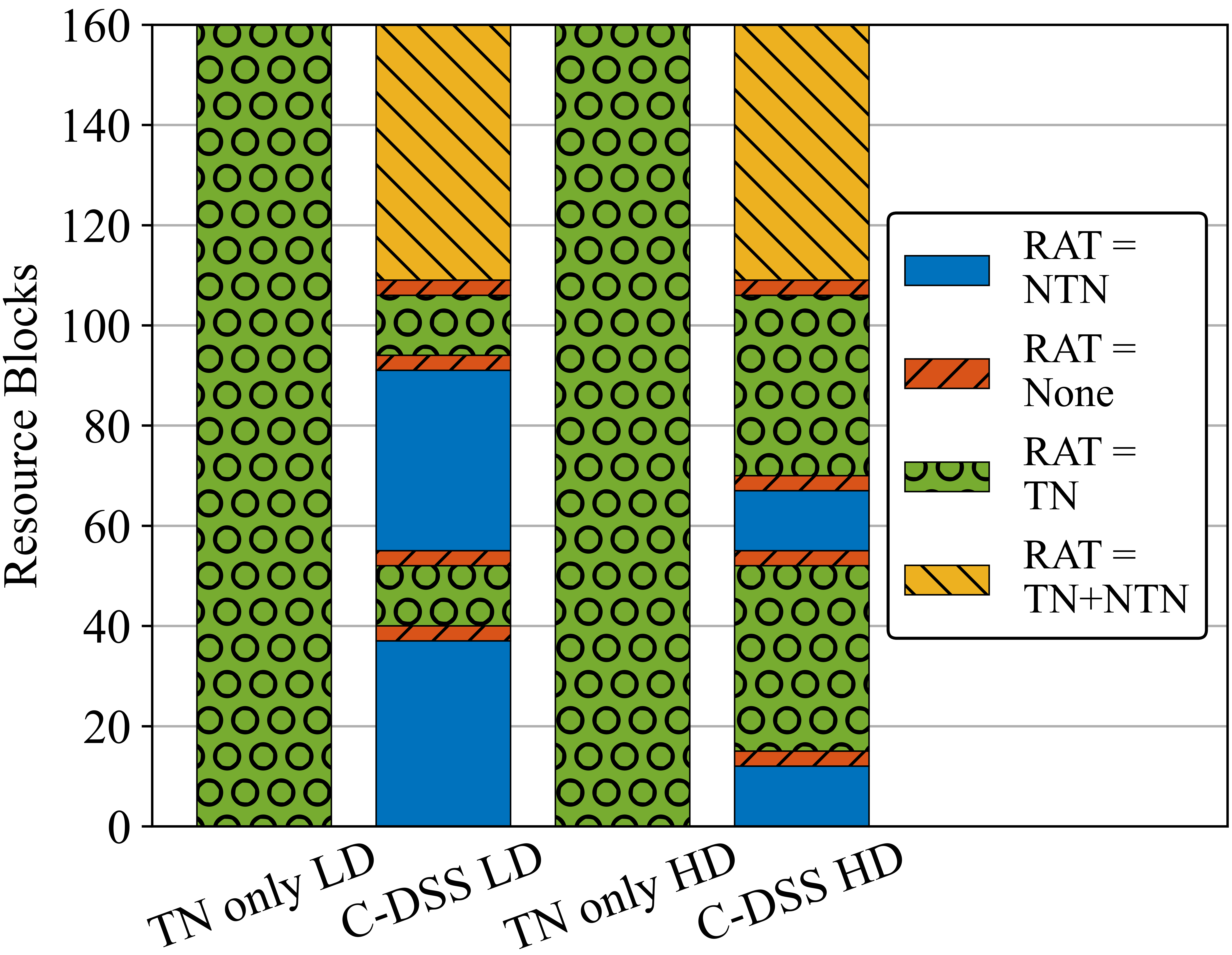}
    \caption{RB allocations at the end of each simulation case.}
    \label{fig:rbs}
\end{figure}

\begin{figure}[htb!]
    \centering
    \includegraphics[width=0.75\linewidth]{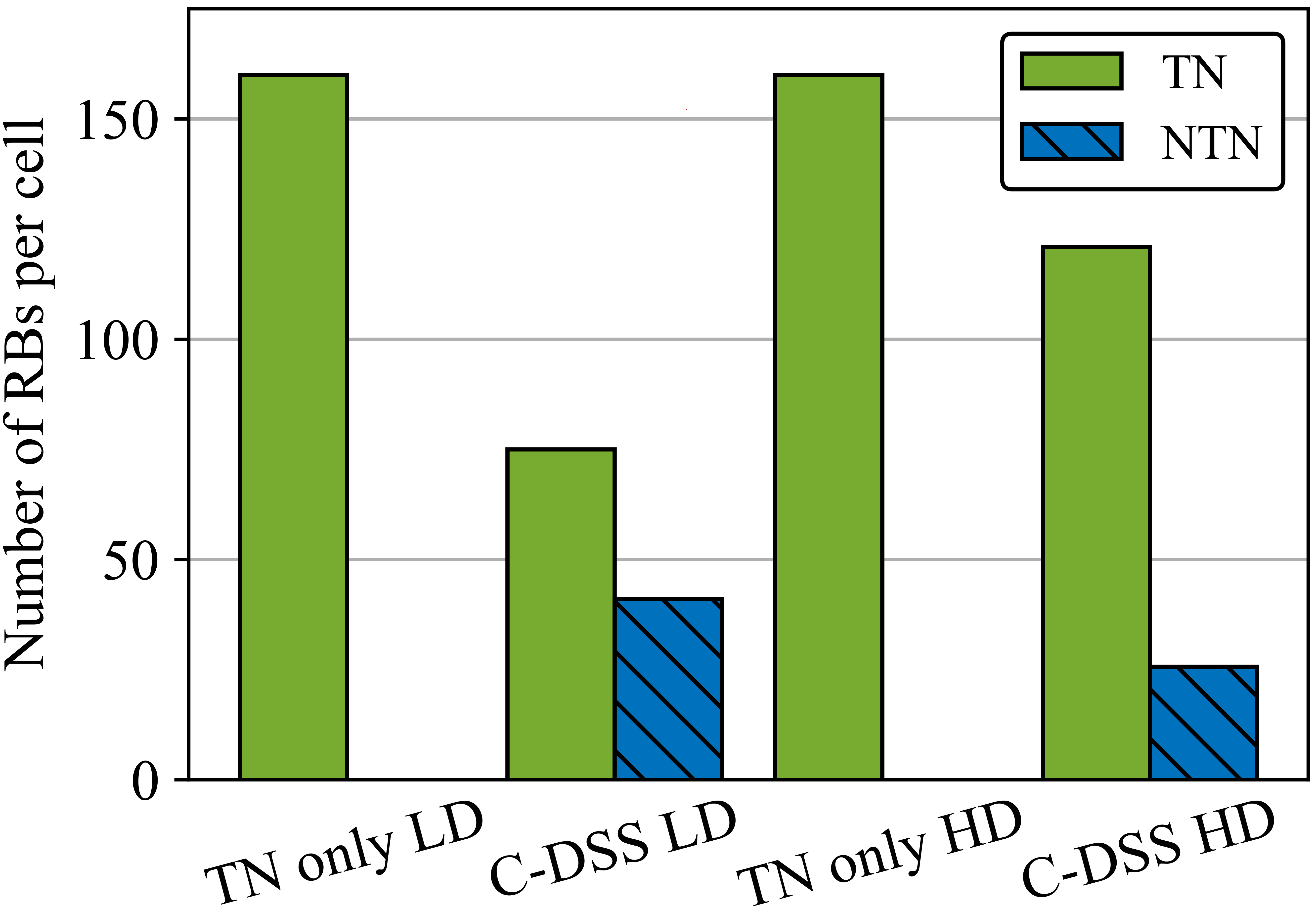}
    \caption{RB counts.}
    \label{fig:RB_counts}
\end{figure}

Fig.~\ref{fig:cdsa_rbs} shows the timeline behavior of case 2 C-DSS with low-demand traffic for different Radio Access Technologies (RAT) combinations, e.g., TN+NTN means both technologies use those particular resources. In the beginning of the simulation, the resources for frequency groups 1 and 2 are shared equally, and group 3 is completely uncoordinated. Due to the low load in TN in this case the NTN beam bandwidth is increased until the TN groups reach the minimum 12 RB allocation. The RB allocations for all simulation cases at the end are shown in Fig~\ref{fig:rbs}. Finally, Fig~\ref{fig:RB_counts} shows the total number of RBs per cell or beam. In cases 1 and 3, the TN gets the full 30 MHz bandwidth which means 160 RBs with selected simulation parameters. In the "C-DSS LD" case more resources are given to the NTN compared to the "C-DSS HD" case. It is notable that in all cases, the TN gets significantly more resources than the NTN due to its FR1 scheme. Out of the 90 TN region users, 25 are connected to the NTN beam 1 or 2 due to the NLOS channel towards the TN gNBs. This means that the total number of users connected to the TN and NTN on average is 65 and 40, respectively.

{}

{}

{}

\begin{figure}[htb!]
    \centering
    \includegraphics[width=0.75\linewidth]{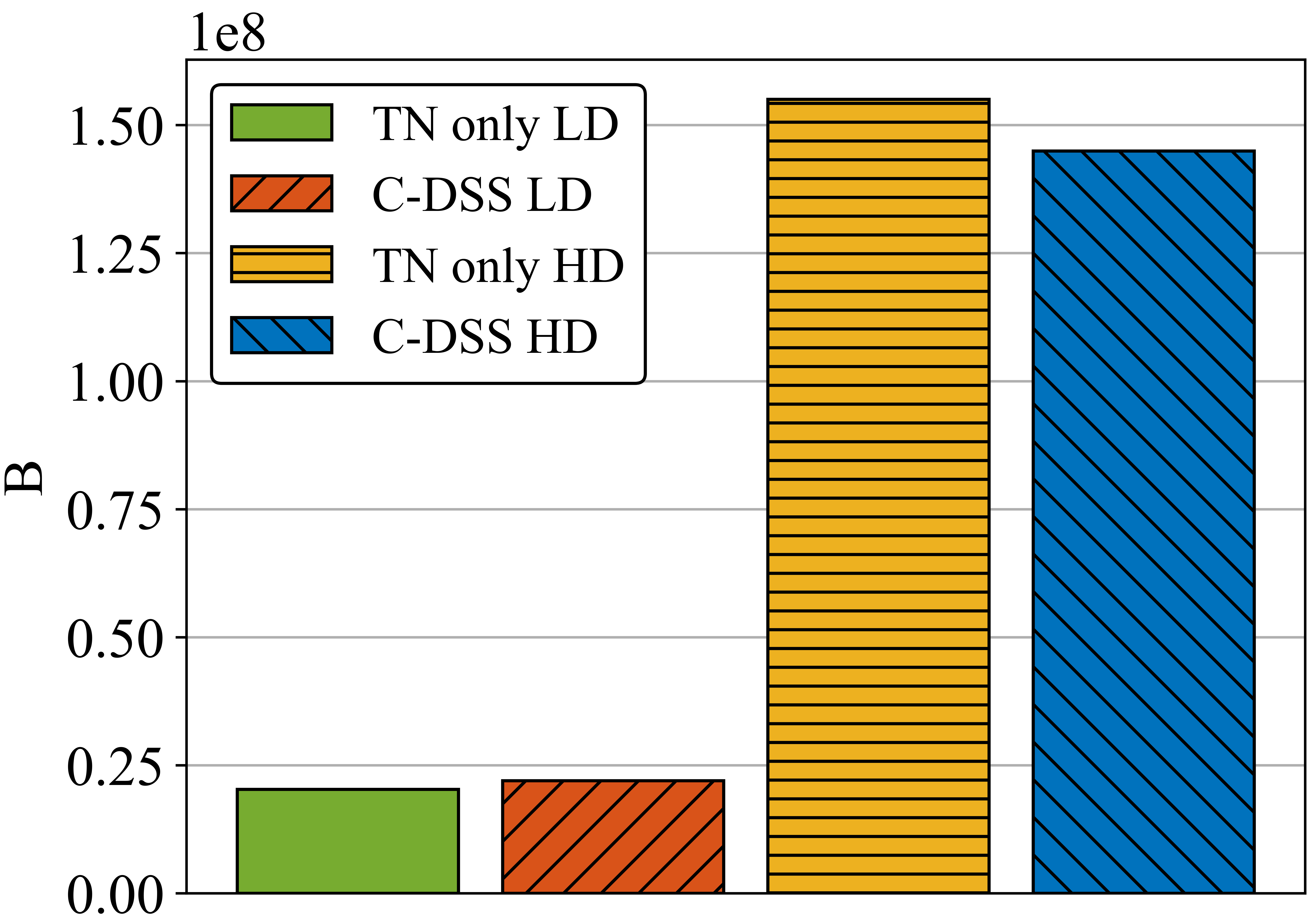}
    \caption{Total RX DL data.}
    \label{fig:rx_dl_data}
\end{figure}

Fig.~\ref{fig:rx_dl_data} shows the total received application bytes during the simulation cases by all users. In low-demand cases 1 and 2, the C-DSS provides the best result although the differences between the cases are small. In these cases, the TN has a low load and there is no need for extra resources and those are better utilized by the NTN users. In high-demand cases 3 and 4 the TN only case gives the highest number as in this case the TN users need the extra resources.

Looking at the throughput distribution in Fig.~\ref{fig:dl_calltp_cdf} we see a different message. In the TN only cases, about 15 \% of users do not get any data service as they are too far away from the TN sites. Most of these users are the ones placed in the NTN beam areas but there are also a few in the TN region. Thus, enabling NTN will significantly improve fairness in the system. Finally, fig.~\ref{fig:dl_calltp_cdf} shows the TN load distribution per cell. There we see how the load varies greatly between the TN cells. Interestingly, C-DSS leads to lower loads with high-demand traffic although getting lower overall throughput. It shows how some of the cells benefit from offloading weak cell edge users to NTN while some TN cells benefit from the extra resources although it means having to potentially serve more users.

{}

\begin{figure}[htb!]
    \centering
    \includegraphics[width=0.75\linewidth]{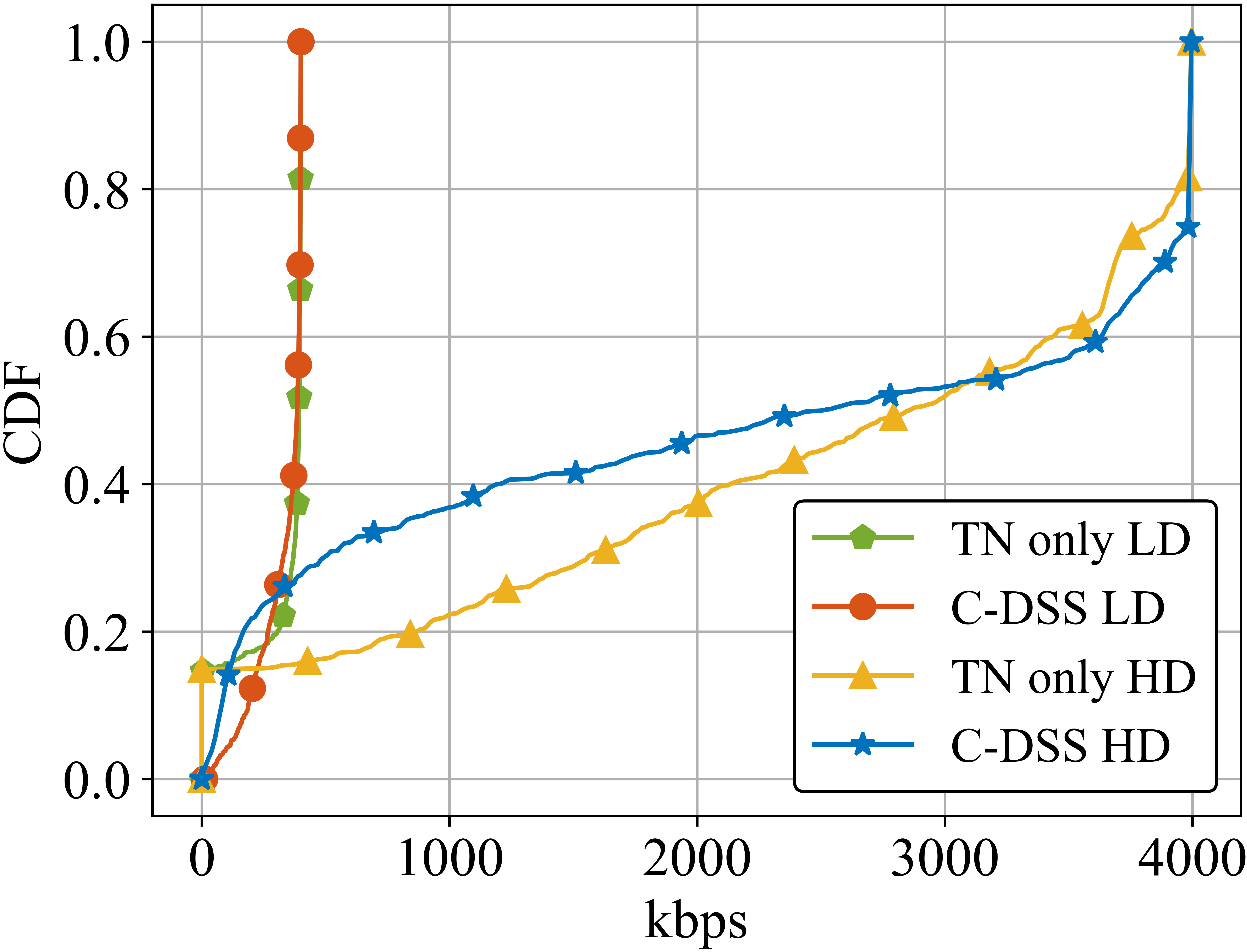}
    \caption{DL call throughput CDF.}
    \label{fig:dl_calltp_cdf}
\end{figure}{}

\begin{figure}[htb!]
    \centering
    \includegraphics[width=0.75\linewidth]{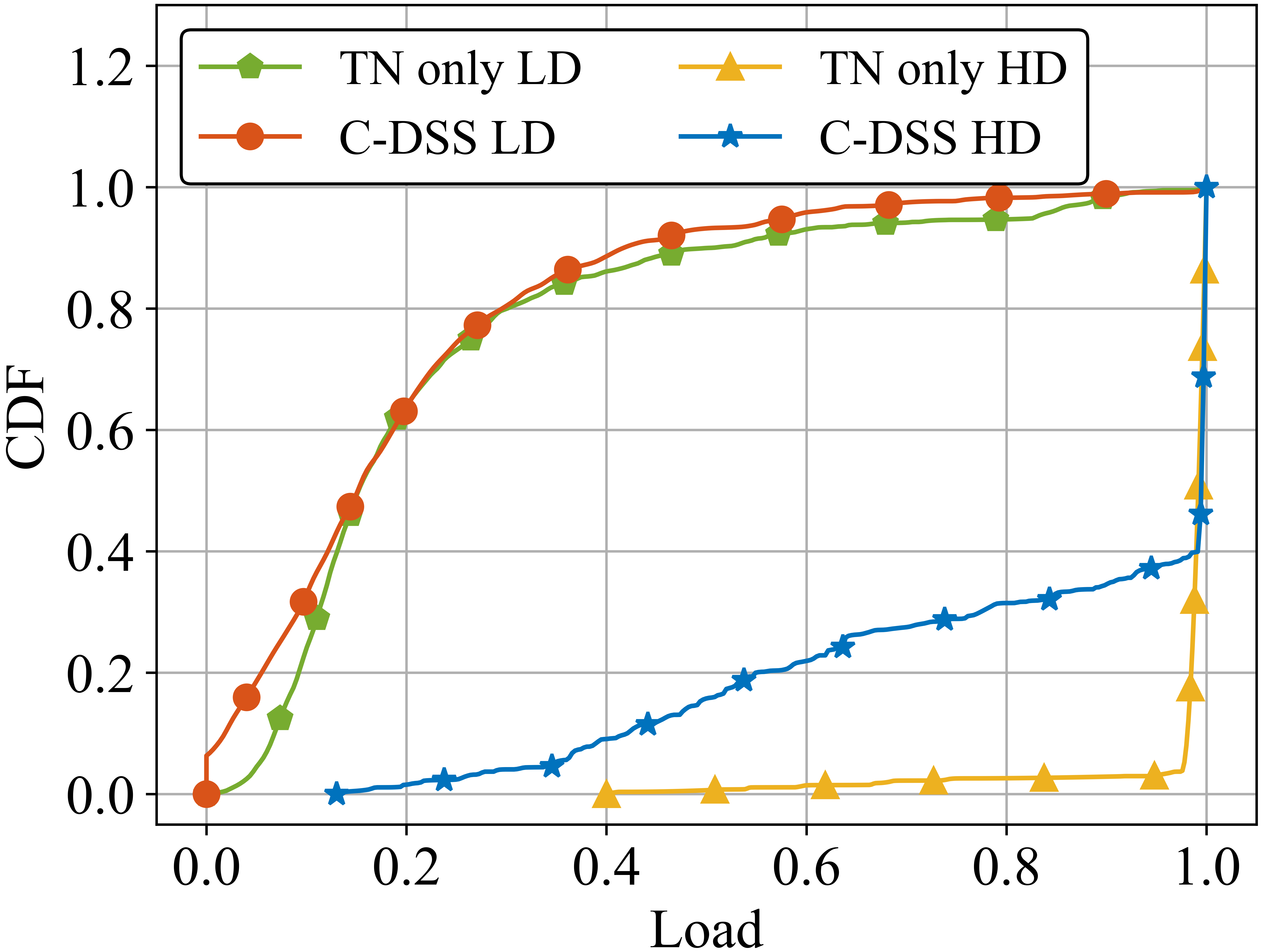}
    \caption{TN DL data resource utilization CDF.}
    \label{fig:dl_calltp_cdf}
\end{figure}{}

\subsection{Result Conclusions}

We have analyzed a scenario where NTN is used to provide coverage to rural area users using a shared bandwidth with a TN operator. The results show how in low traffic demand situations the primary TN users are not affected negatively and NTN can provide service to the rural area. In high-demand traffic situations, the peak performance of the TN inevitably suffers but the TN cell edge and NTN users' performance is improved. The results also show how the load varies from TN cell to cell which limits how much spectrum can be left for NTN. 

\section{Conclusions and Future Work}
\label{sec:conclusions}

In this paper, we have presented a centralized Spectrum Management Server architecture for Coordinated Dynamic Spectrum Sharing (C-DSS) and an algorithm aiming for giving the data resources to the system which needs them while giving priority to TN and taking the peculiarities of the NTN frequency re-use scheme into account. The results show how the C-DSS enables the NTN deployment on a shared spectrum while causing minimal disturbance to primary TN. In the future, more complicated scenarios can be studied which require interference grouping algorithms. Also, scenarios with more dynamic traffic and more sophisticated division of resources between NTN beams should be studied. In addition, further work is needed on how the NR control signaling can operate on limited bandwidth.  

\section*{Acknowledgment}
This work has been funded by the European Union Horizon-2020 Project DYNASAT (Dynamic Spectrum Sharing and Bandwidth-Efficient Techniques for High-Throughput MIMO Satellite Systems) under Grant Agreement 101004145. The views expressed are those of the authors and do not necessarily represent the project. The Commission is not liable for any use that may be made of any of the information contained therein.

\vspace{6pt}
\bibliography{references} 
\bibliographystyle{IEEEtran}

\end{document}